\documentclass[usenatbib,useAMS]{mnras}

\usepackage{graphics,epsf}
\usepackage{amsmath}                
\usepackage{amsfonts}               
\usepackage{amssymb}                
\usepackage{epsfig}                 
\usepackage{graphicx}
\usepackage{comment}

\usepackage{xcolor}
\definecolor{redak}{rgb}{0.9,0.15,0.05}

\def \kms{~\rm{km~s^{-1}}}

\def \msyr{~\rm{M_{\odot}}~\rm{yr^{-1}}}

\def \K{~\rm{K}}

\def \AU{~\rm{au}}

\def \yrs{~\rm{yrs}}

\def \days{~\rm{days}}

\def \rmModot{~\rm{M_{\sun}}}
\def \rmRodot{~\rm{R_{\sun}}}
\def \rmLodot{~\rm{L_{\sun}}}

\title[Accretion in GG Car]{Accretion in the Binary System GG Carinae and Implications for B[e] Supergiants}

\author[A. Kashi]{
Amit Kashi$^{1,2}$\thanks{E-mail: \href{mailto:kashi@ariel.ac.il}{kashi@ariel.ac.il}}
\\
$^{1}$Department of Physics, Ariel University, Ariel, 4070000, Israel\\
$^{2}$Astrophysics, Geophysics and Space Science (AGASS) Center, Ariel University, Ariel, 4070000, Israel\\}

\date{Accepted 9 June 2023. Received 8 June 2023; In original form 19 April 2023}

\pubyear{2023}

\begin{document}
\label{firstpage}
\pagerange{\pageref{firstpage}--\pageref{lastpage}}
\maketitle

\begin{abstract}
We simulate the hydrodynamics of the wind flow in the B[e] supergiant binary system GG~Carinae and obtain the mass accretion rate onto the secondary and the observed lightcurve. We find an inhomogeneous Bondi-Hoyle-Lyttleton accretion into a curved accretion tail, and confirm that the accretion rate is modulated along the orbit, with a maximum close to periastron.
We show that the accretion itself cannot account for the periodical variation in brightness. Instead, we explain the observed variation in the light curve  with absorption by the accretion tail.
Our results are in general agreement with previously derived stellar masses, orbital parameters, and the system orientation, but imply that the B[e] supergiant is more luminous. 
We find an effect related to the orbital motion of the two stars, in which the accretion tail is cut by the primary and we term it the Lizard Autotomy Effect. As part of the effect, the primary is self accreting wind that it ejected earlier.
The Lizard Autotomy Effect creates an outwardly expanding spiral shell made up of broken segments.
We suggest that such a tail exists in other B[e] supergiant systems and can be the source of the circumstellar material observed in such systems.
The accretion also forms a disc around the secondary near periastron that later vanishes. We suggest that the formation of such a disc can launch jets that account for the bipolar structure observed around some B[e] supergiants.
\end{abstract}


\section{INTRODUCTION}
\label{sec:intro}

Massive star evolution is an intriguing process that includes stages that are not yet fully understood \citep[e.g.,][]{Smartt2009, Owocki2011, DavidsonHumphreys2012, Vink2015, Georgyetal2017, Owockietal2019, Farrelletal2022, EldridgeStanway2022}.
Amongst other massive stars at evolved stages, the B[e] supergiant stars (B[e]SGs or sgB[e]) are rare stars, and only a few tens of them are known. 
B[e]SGs exhibits optical spectra with strong Balmer emission lines, low excitation permitted emission lines of predominantly low ionization metals, and forbidden emission lines of [Fe~II], [O~I] and other species, as well as strong infrared excess in their spectrum, strong continua and few if any apparent absorption lines
\citep[e.g.][]{AllenSwings1976,Zickgrafetal1986,Zickgrafetal1996,Lamersetal1998, Miroshnichenko2006,Zickgraf2006,Aretetal2016,Humphreysetal2017, OudmaijerMiroshnichenko2017,Kraus2019}.

As such, and having approximately the same effective temperature but lower luminosities ranging between $L=10^4$--$10^6 \rmLodot$, B[e]SGs may be related as the fainter, lower mass siblings of Luminous Blue Variables (LBVs) \citep[e.g.,][]{Zickgrafetal1996, Clarketal2013, Krausetal2014}. As the luminosity ranges overlap it may be difficult to distinguish between them, with the main way being the lack of [O~I] emission in LBVs \citep{Humphreysetal2017}.
Lower mass stars sometimes show similar properties to the B[e] phenomenon, but not being SGs they do not belong to the group, and can be classified as FS~CMa stars \citep[e.g.,][]{Miroshnichenko2006, Korcakovaetal2022, Miroshnichenkoetal2023} and might be close binaries that are undergoing or have
undergone in the past a phase of mass transfer between the components (see also \citealt{Nodyarovetal2022} for a recent example of such a system). Some yellow hypergiants might also appear as B[e]SGs when passing the yellow void in the HR diagram \citep[e.g.,][]{Daviesetal2007,Aretetal2017a,Aretetal2017b}.

B[e]SGs are surrounded with a circumstellar molecular and dusty disc or ring, but the mechanism of how this circumstellar material arranged into its shape and structure is not clear \citep{Conti1997, Lamersetal1998, Krausetal2013, Kraus2019}.
The most stable and most abundant molecule emitting in the near-IR in the discs around B[e]SGs is CO. The spectrum also shows emission from the isotopic molecule $^{13}$CO, indicating that the circumstellar material has been released from the evolved star \citep{Kraus2009, Krausetal2022}.
The rarity of B[e]SGs may indicate that they are a fast transitional phase in massive stars evolution of a certain mass range, characterized by strong mass loss. A further strengthening to this conjecture comes from the fact that many B[e]SGs show not only small-scale molecular and dusty rings, but also large-scale nebulae and ejecta, suggesting previous mass-loss events \citep[e.g.,][]{Krausetal2021}.

B[e] stars have been found to also have large-scale (up to several pc) gas with various shapes: spherical and ring nebulae, spiral-arms or partial spiral arms, bipolar and unipolar-lobe structures \citep[e.g.,][]{MarstonMcCollum2008}.
Radial velocities (RV) have been measured in the circumbinary material around B[e] stars, suggesting an outflowing velocity component \citep{Krausetal2010}.
Observations of the spatial distribution of the circumstellar gas revealed that it is more likely accumulated in multiple rings or arcs, and possibly in spiral arm-like structures \citep{Torresetal2018,Krausetal2016, Maraveliasetal2018, Liimetsetal2022}.

Some B[e]SGs were suggested to be in binary systems, and in this case the circumstellar disc is in fact a circumbinary disc.
According to \cite{Georgyetal2017} it is difficult to explain the asymmetries in the circumstellar medium of B[e]SGs with a single star, since their stellar evolution models predict small rotation rates for blue supergiants at the observed position of B[e]SGs in the H-R diagram.
Thus, the similarity between the B[e]SGs, LBVs and FS~Cma might support a binary origin for the B[e]SGs as well.

GG~Carinae (HD 94878) is one of these confirmed binary B[e]SG systems, that is relatively well observed \citep{Swingsetal1974,Hernandezetal1981,Gossetetal1985,Brandietal1987,Machadoetal2004,Pereyraetal2009,Marchianoetal2012,Krausetal2013,Porteretal2021,Porteretal2022}.
GG~Car is considered to be the archetypal B[e]SG system due to the high B[e]SG mass $M_1=24\pm4 \rmModot$, luminosity $\log(L_1/\rmLodot)=5.26 \pm 0.19$ \citep{Porteretal2021}, and complex circumstellar environment. However, its peculiar behavior on both short and orbital time scales suggests that it is an atypical system.
Most previous evolutionary studies of GG~Car assumed that the two binary components have evolved as pseudo-single stars \citep{Kraus2009, Kraus2019, Krausetal2013, Oksalaetal2013}.
\cite{Porteretal2021} brought a different view, modeling the system as a binary with interaction between its components.

According to radiation transfer calculations with the \textsc{cmfgen} code computed by \cite{Porteretal2021}, the observed spectral behavior of GG~Car implies that the emission lines are generated by the primary's wind, and the varying amplitudes of the different line species are attributed to their formation at different radii, with less energetic lines forming at larger radii. 
They also determined the binary orbit by modeling the atmosphere of the primary and the radial velocity variations of the wind lines.
They concluded that the period, which was earlier unclear, is $P=31.01\pm0.01\days$ and the eccentricity is $e = 0.50 \pm 0.03$, much larger than earlier estimates \citep{Marchianoetal2012,Krausetal2013}.
From their fit to the orbital solution \cite{Porteretal2021} also found that the argument of periastron is $\omega=339.87^{\circ} \pm 3.1^{\circ}$.
\cite{GrantBlundell2022} used the marginalized GP algorithm and found similar orbital eccentricity and $\omega=334.36^{\circ} \pm 6.1^{\circ}$.
The inclination of the system is $i\approx60^{\circ}$ \citep{Krausetal2013}, but it is not very well constrained, perhaps to within $20^{\circ}$ \citep{Porteretal2021}.

To explain the variation in luminosity along the orbit,
\cite{Porteretal2021} suggested that the system undergoes enhanced \textit{mass transfer and accretion} from the primary to the secondary at periastron.
While the primary component of GG~Car is well identified, the class of the secondary component is still unknown, though it is much less luminous than the primary. \cite{Porteretal2021} obtained a preferred set of parameters for both stars in the system as well as the binary orbit (values cited below).

The primary wind parameters obtained by \cite{Porteretal2021} are unusual as the mass loss is relatively low and the wind is slow. Massive stars with similar mass loss rate usually have higher wind velocities \citep[e.g.][]{Pulsetal2008,Owockietal2015,VinkSander2021}. We therefore test another set of parameters based on the Geneva stellar evolution code \citep{Ekstrometal2012}. The simulations results that derived from using the two sets are then compared to observations.

The GG~Car system has a very weak x-ray emission, indicating that it is not a colliding-wind binary such as $\eta$ Car \citep{PittardCorcoran2002,Akashietal2006,KashiSoker2009a,Kashietal2021} or a few other LBVs \citep[e.g.,][]{Nazeetal2012}. Namely, the secondary is in an evolutionary state in which its wind is very weak compared to the primary and has a low wind momentum ratio. This indicates that the accretion is expected to be Bondi-Hoyle-Lyttleton (BHL), or a fraction of this rate \citep[e.g.,][]{Kashietal2022}.

The spectroscopic study of \citep{Porteretal2022} found that GG~Car has two circumbinary rings. The rings are orbiting with projected velocities of 84.6 and 27.3 $\kms$ at radii of $2.8^{+0.9}_{-1.1} \AU$ and $27^{+9}_{-10} \AU$ for the inner and outer rings, respectively. They further found variations in the rings consistent with pumping of the eccentricity of a radially thin circumbinary ring by the inner binary.
The ring or rings are far from the binary and its immediate vicinity which are the focus of our paper, but we will discuss how they might have formed.

In this paper we a run numerical simulations for the GG~Car binary system obtaining the geometry of the wind as a result of binary interaction as well as accretion onto the secondary.
In section \ref{sec:simulation} we describe the numerical simulation.
The results from our two sets of stellar and wind parameters are presented in section \ref{sec:results}. We discuss our results in section \ref{sec:discussion}.
Our summary and conclusions are given in section \ref{sec:summary}.

\section{THE NUMERICAL SIMULATIONS}
 \label{sec:simulation}

\begin{figure*}
\centering
\includegraphics[trim= 0.0cm 0.0cm 0.0cm 0.0cm,clip=true,width=0.50\textwidth]{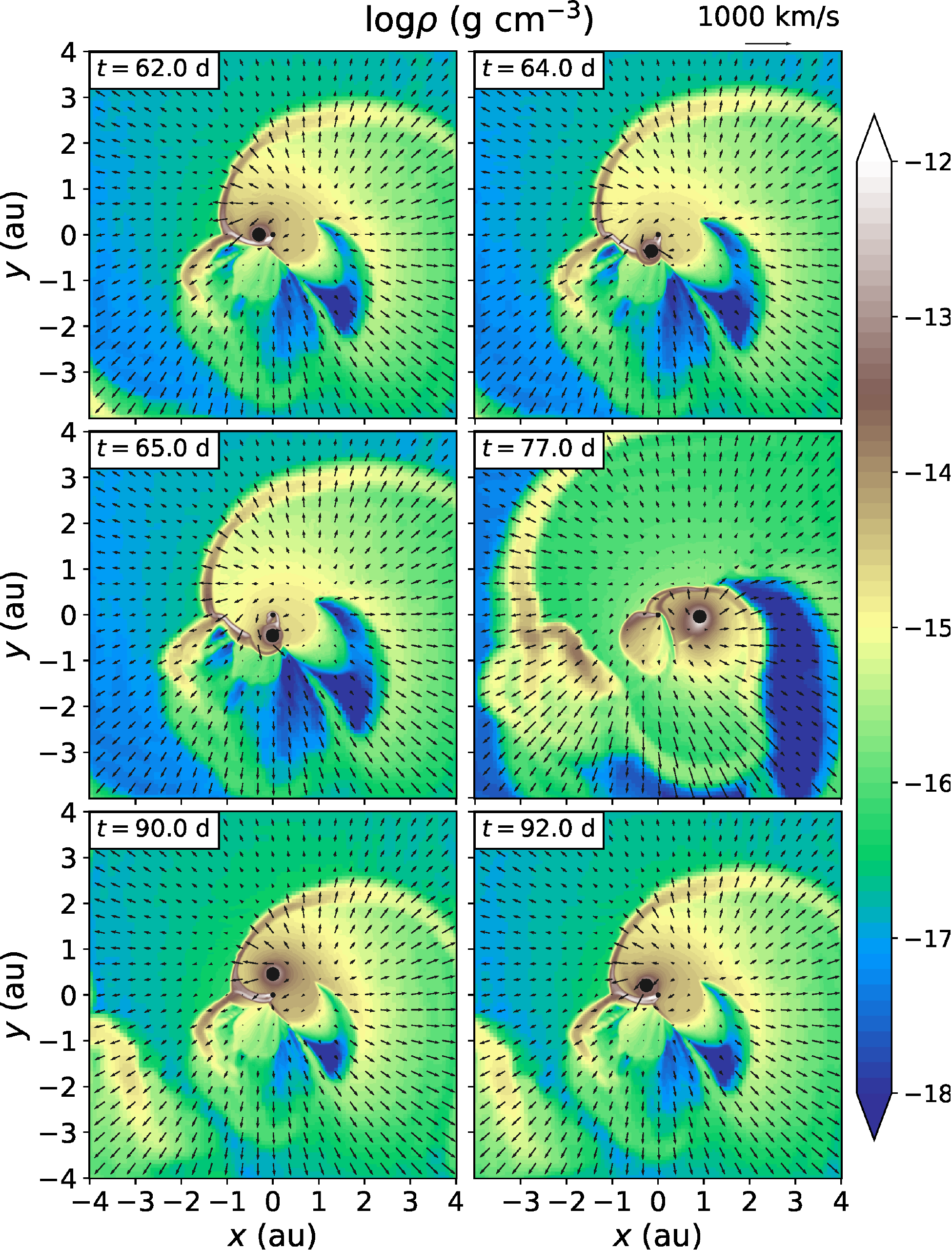} 
\includegraphics[trim= 0.0cm 0.0cm 2.1cm 0.0cm,clip=true,width=0.46\textwidth]{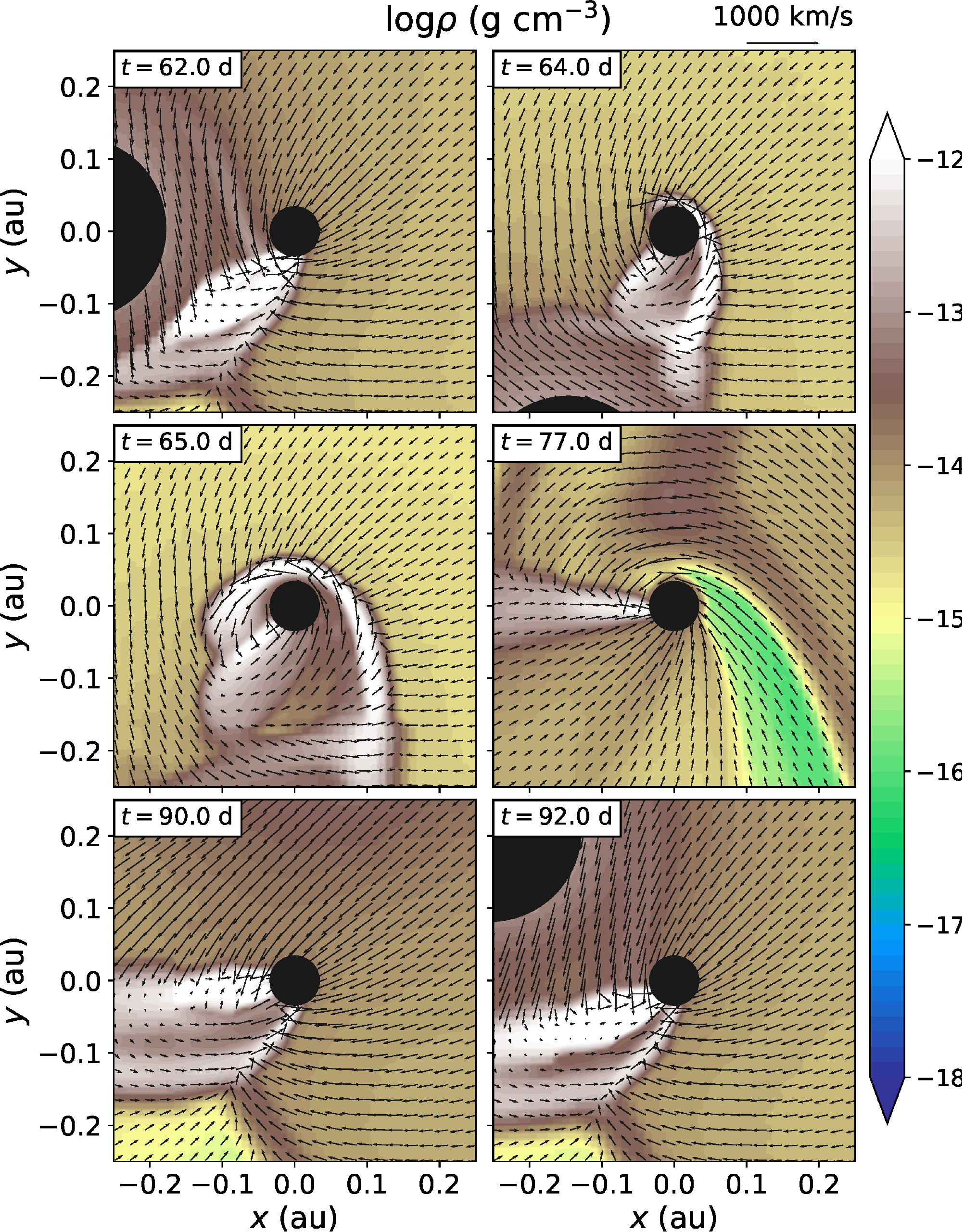} 
\caption{
\textit{Left:} Density maps with velocity vectors showing slices in the orbital plane ($z=0$), at different times for set P.
The times start after two orbits in which we let the system relax and shows the third orbit from periastron ($t=62 \days$), through apastron (middle right, $t=77 \days$) till one day before the next periastron (lower right).
The periastron passage takes place when the primary is at $(-0.305, 0, 0)\AU$, and moving counterclockwise when looking from `above' the grid (positive $z$).
The secondary is at the center, with its radius indicated by a small black circle. The primary, with its radius indicated by a large black circle, orbits the secondary counterclockwise, with periastron occurring when the primary is left of the secondary.
The accretion tail is obtained and funnels the primary wind to be accreted onto the secondary.
Close to periastron passage ($t=62 \days$) the tail is being cut by the passing of the secondary close to the primary.
\textit{Right:} Focusing on the secondary at the same times as on the panels on the left hand side. Note that the color scheme is the same for both panels but the velocity scale is different, and more velocity vectors are plotted for the focus panels. An accretion disc-like structure around the secondary is obtained shortly after periastron passage. At later times the accretion is inhomogeneous BHL.
}
\label{fig:dens_6panel}
\end{figure*}

\begin{figure*}
\centering
\includegraphics[trim= 0.0cm 0.0cm 0.0cm 0.0cm,clip=true,width=0.50\textwidth]{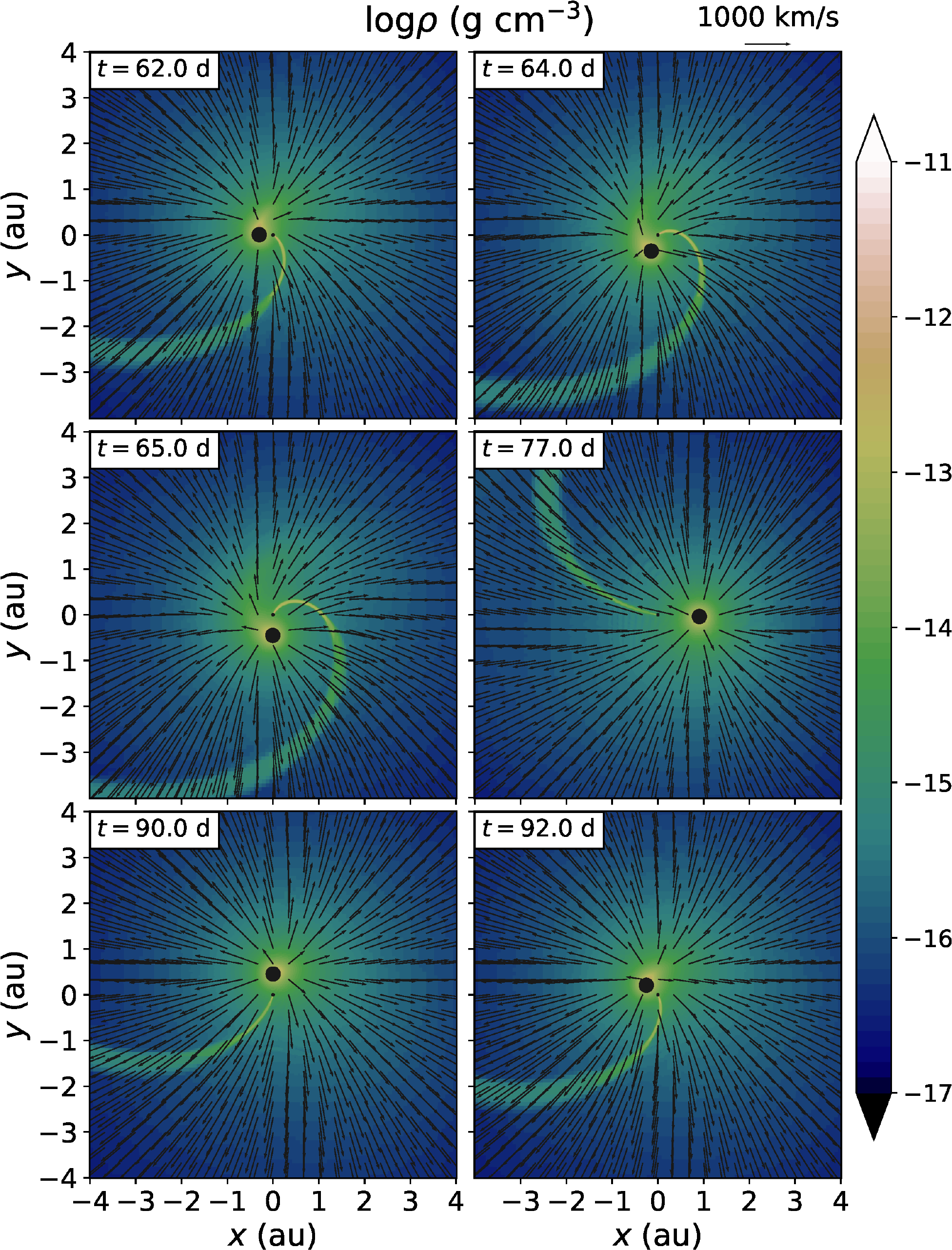} 
\includegraphics[trim= 0.0cm 0.0cm 2.1cm 0.0cm,clip=true,width=0.47\textwidth]{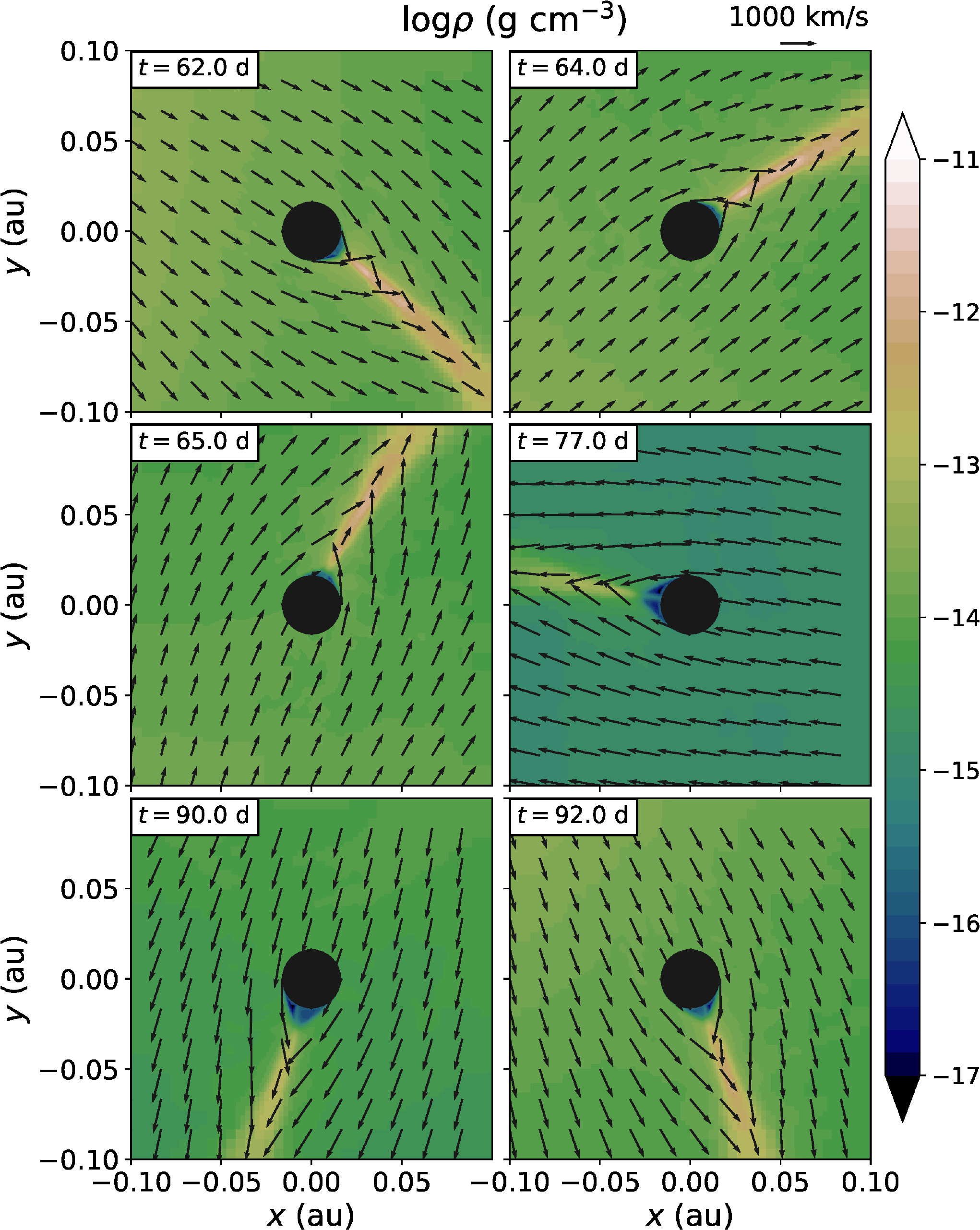} 
\caption{
Same as Figure \ref{fig:dens_6panel}, but for set G. The terminal velocity of the primary for this set $v_{\rm 1,inf}=1400 \kms$ is much larger than the orbital velocity $v_{\rm orb,p}<<v_{\rm 1,inf}$,
Therefore a curved tail that changes its direction along the orbits obtained. Differently from model P, the primary does not cut the tail as in set P and the Lizard Autotomy Effect does not take place.
}
\label{fig:dens_6panel_set_G}
\end{figure*}

We run the numerical simulation using the hydrodynamic code \textsc{flash}, described in \cite{Fryxell2000}.
Our 3D Cartesian grid extends over $x\in(-4,4) \AU$, $y\in(-4,4) \AU$ and $z\in(-2,2) \AU$. The grid is centered around the secondary star, and the orbital plane is on the $x$--$y$ plane at $z=0$.
To solve the hydrodynamic equations we use the \textsc{flash} version of the split piece-wise parabolic method (PPM) solver \citep{ColellaWoodward1984} which is an extension of the Godunov method, that solves a Riemann problem and allows resolving shocks.
A multigroup diffusion approximation is used for solving the radiative transfer equation and updating the energy.
We also include radiative cooling from \cite{SutherlandDopita1993}, which is necessary in modeling colliding winds governed by instabilities. The temperature is limited from below to $1\,000\K$.

We set up the GG~Car binary system using two sets of stellar parameters. The first set, designated `P', is based on the parameters from in \cite{Porteretal2021}, noting that some of them were derived in earlier papers. The binary semi-major axis is $a=0.61 \AU$, the orbital period is $P=31.01$ days and the eccentricity is $e=0.5$. The stellar masses of the primary B[e]SG star and its companion are ${M_1 = 24 \rmModot}$ and ${M_2=7.2\rmModot}$, respectively. We note that for the masses the range in the literature was quite large. The effective temperature of the primary is $T_{\rm eff,1} = 23\,000 \K$, its radius is taken to be $R_1= 27 \rmRodot$ and the luminosity is $L_1 \simeq 1.8 \times 10^5 \rmLodot$. The secondary is hotter and smaller with $T_{\rm eff,2} = 16\,000 \K$, $R_2= 7.2 \rmRodot$, and $L_2 \simeq 3\,000 \rmLodot$.
The composition used for both stars is appropriate for an evolved star with CNO-processed material and the He abundance is taken to be 50 percent, and the metalicity is $Z=0.02$. This composition is also used for calculating the opacity, for which an \textsc{opal} model is used \citep{IglesiasRogers1996} and low temperature opacities are also included.
The interiors of the stars are excluded from the simulation. The gravitational field is modeled by two point sources at the centers of both stars.
The winds are ejected from the grid cells in a narrow spherical shell around each star. Both stars accelerate their winds according to the $\beta$ velocity law $v(r)=v_{\infty} \left(1- R_\ast/r \right)^{\beta}$, where $v_{\infty}$ is the terminal velocity, $R_\ast$ is the stellar radius, $r$ is the radial distance measured from the center of the star and $\beta$ is the wind acceleration parameter.
The primary mass loss rate is $\dot{M}_1 = 2.2 \times 10^{-6} \msyr$ and its wind velocity has a terminal value of $v_{1,\infty}=265 \kms$ with radiative acceleration corresponding to $\beta=1$.
The secondary mass loss rate is $\dot{M}_2 =10^{-8} \msyr$, and its wind velocity has a terminal value $v_{2,\infty} = 1\,600 \kms$ with an acceleration parameter $\beta=1$.
The grid has 6 levels of refinement that are kept constant with time.
The largest cell has a side of $\simeq 18 \rmRodot$ and the smallest has a side of $\simeq 0.6 \rmRodot$. 
The periastron passage takes place when the primary is at $(-0.305, 0, 0)\AU$, and moving counterclockwise when looking from `above' the grid (positive $z$).

The second set of parameters we use, designated `G', is based on a stellar track from the Geneva code \citep{Ekstrometal2012}.
Using an evolutionary track of a rotating star with zero age main sequence mass $M_{1,\rm ZAMS}=25\rmModot$ for the primary, we set our simulation at $M_{1}\simeq23.6\rmModot$, when the effective temperature is $T_{\rm 1,eff} = 23\,800 \K$, closest to the value in model P.
The mass loss rate is $\dot{M}_1 \simeq 4 \times 10^{-6} \msyr$, and the terminal velocity we derive  is $v_{1,\infty} =2 \simeq 1400\kms$. 
The main difference between the two sets of parameters is therefore that $v_{1,\infty}$ is much larger for set G.
For the secondary we use $M_{2,\rm ZAMS}=7.2\rmModot$ which gives almost the same mass at the time we take for the primary, as it is still a main-sequence star, $M_{2}\simeq7.2\rmModot$. The terminal wind velocity is $v_{2,\infty} \simeq 900\kms$ and the mass loss rate is $\dot{M}_2 \simeq 6.24 \times 10^{-12} \msyr$.
Table \ref{table:stellarandorbitalparameters} gives the full list of parameters in both models.
The orbital parameters of set G are similar to those of set P (except for a minor difference of $<0.7$\% in the semi-major axis).
As the secondary is smaller, for set G the grid has 8 levels of refinement, and the smallest has a side of $\simeq 0.14 \rmRodot$. 

\begin{table}
\centering
\caption{
List of stellar, wind and and orbital parameters for the two sets of parameters used in our simulations. Set P is based on parameters from Porter et al. (2021). Set G is based on a stellar track from the Geneva code.
}
\begin{tabular}{lll} 
\hline
Parameter                       & Set P                   & Set G                   \\
\hline
$M_1 (\rmModot)$                & $24$                    & $23.6$                  \\
$R_1 (\rmRodot)$                & $27$                    & $31$                    \\
$T_{\rm eff,1} (\K)$            & $23\,000$               & $23\,800$               \\
$L_1 (\rmLodot)$                & $1.8 \times 10^5 $      & $2.4 \times 10^5 $      \\
$\dot{M}_1 (\msyr)$             & $2.2 \times 10^{-6}$    & $4 \times 10^{-6}$      \\
$v_{1,\infty} (\kms)$           & $265$                   & $1\,400$                \\
$\beta_1$                       & $1$                     & $1$                     \\

\hline

$M_2 (\rmModot)$                & $7.2$                   & $7.2$                  \\
$R_2 (\rmModot)$                & $7.2$                   & $3.4$                  \\
$T_{\rm eff,2} (\K)$            & $16\,000$               & $20\,800$              \\
$\dot{M}_2 (\msyr)$             & $ 10^{-8}$              & $6.24 \times 10^{-12}$ \\
$v_{2,\infty} (\kms)$           & $1\,600 $               & $900 $                 \\
$\beta_2$                       & $1$                     & $1$                    \\
\hline
$P$ (days)                      & $31.01$                 & $31.01$                \\
$e$                             & $0.5$                   & $0.5$                  \\
$a (\AU)$                       & $0.61$                  & $0.61$                 \\ 

\hline
&&\\
\end{tabular}
\label{table:stellarandorbitalparameters}
\end{table}

For both sets of parameter, the wind parameters of the secondary have a very small impact on the flow structure of the simulation as its wind momentum is negligible compared to that of the primary. So the grid is dominated by almost exclusively the primary wind, that is shaped by the orbital motion and the gravity of the secondary.
We ran one simulation for each set of parameters with one set of parameters, but post-process it for four different orientations.

Our simulation code does not assume that accretion takes place. We design the code to allow treatment of accretion \textit{if} material flows towards the secondary and not stopped by  radiation braking. The accretion treatment includes measuring the accreted material, its arrival direction and other properties.
We quantify the accretion rate according to the method described in \citep{Kashi2019}. The secondary is resolved as a sphere (not a sink) and accretion is calculated in each cell around the secondary. 

\section{RESULTS}
\label{sec:results}

\subsection{Set P}
\label{sec:resultsP}

\begin{figure*}
\centering
\includegraphics[trim= 1.0cm 2.0cm 3.0cm 2.0cm,clip=true,width=1.0\textwidth]{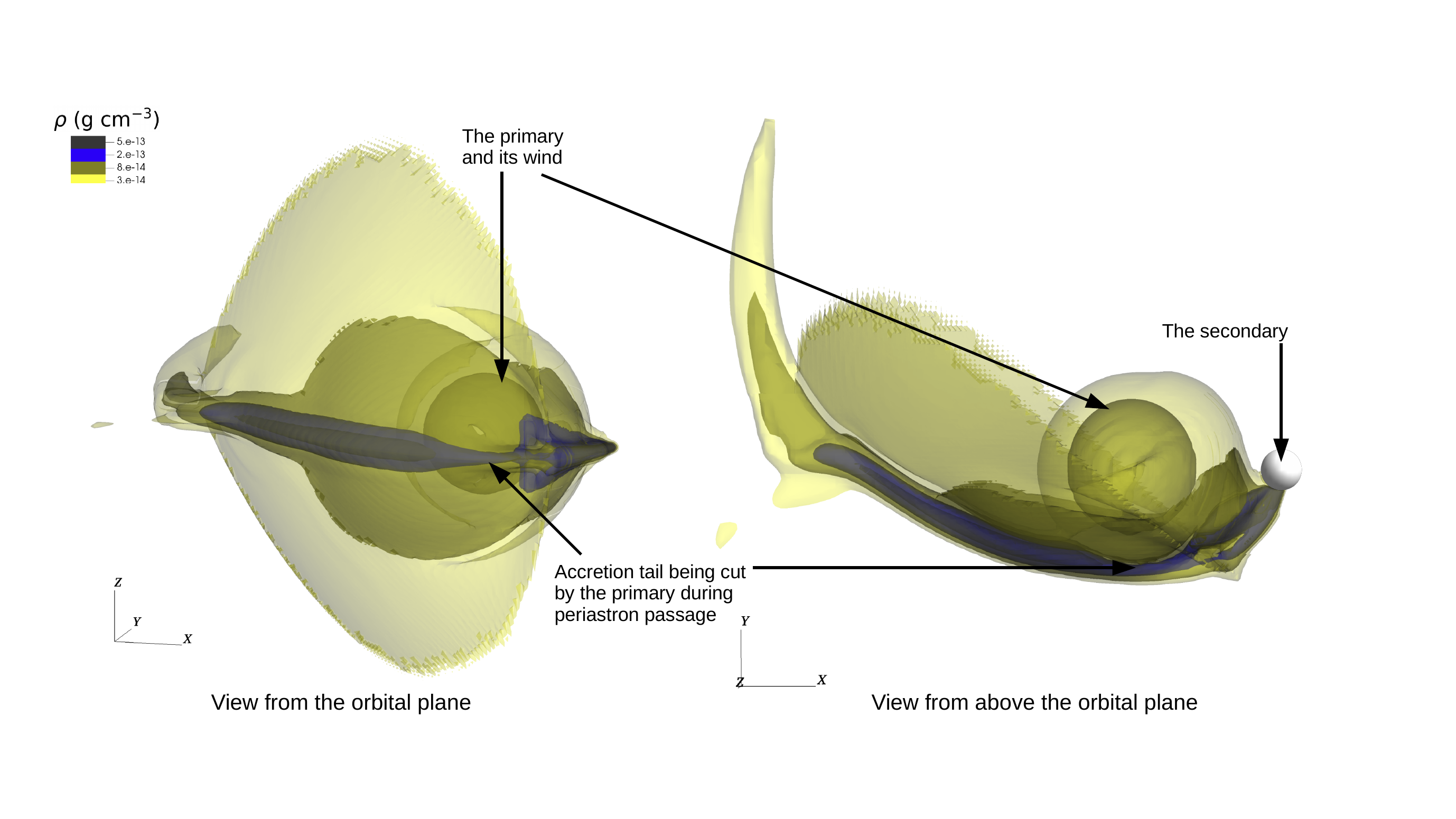}
\caption{
A 3D view of the simulation with parameters set P. The accretion tail being cut by the primary during periastron passage, shown from two viewing angles. The secondary does not appear in the 3D view, and was added for perspective and to show the scale ($R_2=7.2\rmRodot$).
There is a dense tail behind the secondary, part of which is an accretion tail towards the secondary and part is outflowing away from it.
The secondary takes a sharp elliptical turn close to periastron and as a result the close primary and its wind cut the tail and creates a gap in the tail -- the Lizard Autotomy Effect.
The process repeats every periastron passage, creating an outflowing spiral with gaps.
}
\label{fig:3d}
\end{figure*}

We start with set P. The simulation outputs are exported and post-processed every 0.5 day. 
Figure \ref{fig:dens_6panel} shows a slice of the simulation grid on the orbital plane.
The right six panels are zoom-in on the secondary at times corresponding to the left six panels, and have the same density scale, but more velocity vectors.
The times given in Figure \ref{fig:dens_6panel} in the upper left of each panel are during the third orbit, as for two orbital periods we let the system relax from the initial condition that did not include wind from previous orbits but only wind that started to flow from the two stars.
Therefore after two orbits the wind structure is representative of the system unlike the initialization of the simulation.
The times show the third orbit from periastron (upper left panel in each six panel group), through apastron (middle right) till one day before the next periastron (lower right).

As expected, we find that the accretion takes a geometric form that resembles inhomogeneous BHL accretion \citep{Livioetal1986}. The primary wind flows from both sides of the secondary and focused by its gravity into a curved tail. Part of the accretion tail flows towards the secondary and gets accreted, and part of it flows outwards and creates a spiral structure.
Accretion also takes place from other directions but at a lower rate.
After periastron passage a clear dense accretion disc is obtained around the secondary (see the middle left panel at $t=65.0 \days$). The disc lasts for a few more days and dissipates as the system approaches apastron (middle right panel at $t=77.0 \days$).
An accretion tail remains even after the dissipation of the disc, and lasts until the next periastron passage.
The flow creates a spiral, with higher density close to the secondary.

Figure \ref{fig:3d} shows a 3D contour plot of the high density parts of the simulation close to periastron passage.
These dense parts include the strong wind of the primary and the  tail of the secondary.
There is a dense tail behind the secondary, part of which is an accretion tail towards the secondary and part is outflowing away from it.
We can see the secondary (added artificially on the right side that views the system from above the orbital plane) close to periastron passage.
As the secondary takes a sharp elliptical turn when coming to periastron and passes close to the primary and its wind, the tail is being cut by the primary and a gap in the tail is created.
As this process is reminiscent of the self cutting of a lizard's tail we will refer to it as the \textit{Lizard Autotomy Effect}.
The process repeats every periastron passage, creating an outflowing spiral with gaps, that also has a rotating velocity component.

Another effect that we obtain as part of the Lizard Autotomy Effect is \textit{self accretion}: the primary passes through the tail and accretes its own wind.
We did not measure quantitatively the accreted mass onto the primary, as our code is not designed to do so (it only measures the mass accretion onto the secondary).
The self accreted material comes mostly from one direction -- the side of the primary facing the tail. As the density of the tail is much larger than the density of the direct primary wind (Figure \ref{fig:3d}) it is unlikely that the latter can prevent the accretion.

\begin{figure}
\centering
\includegraphics[trim= 1.2cm 0.0cm 0.0cm 0.0cm,clip=true,width=0.99\columnwidth]{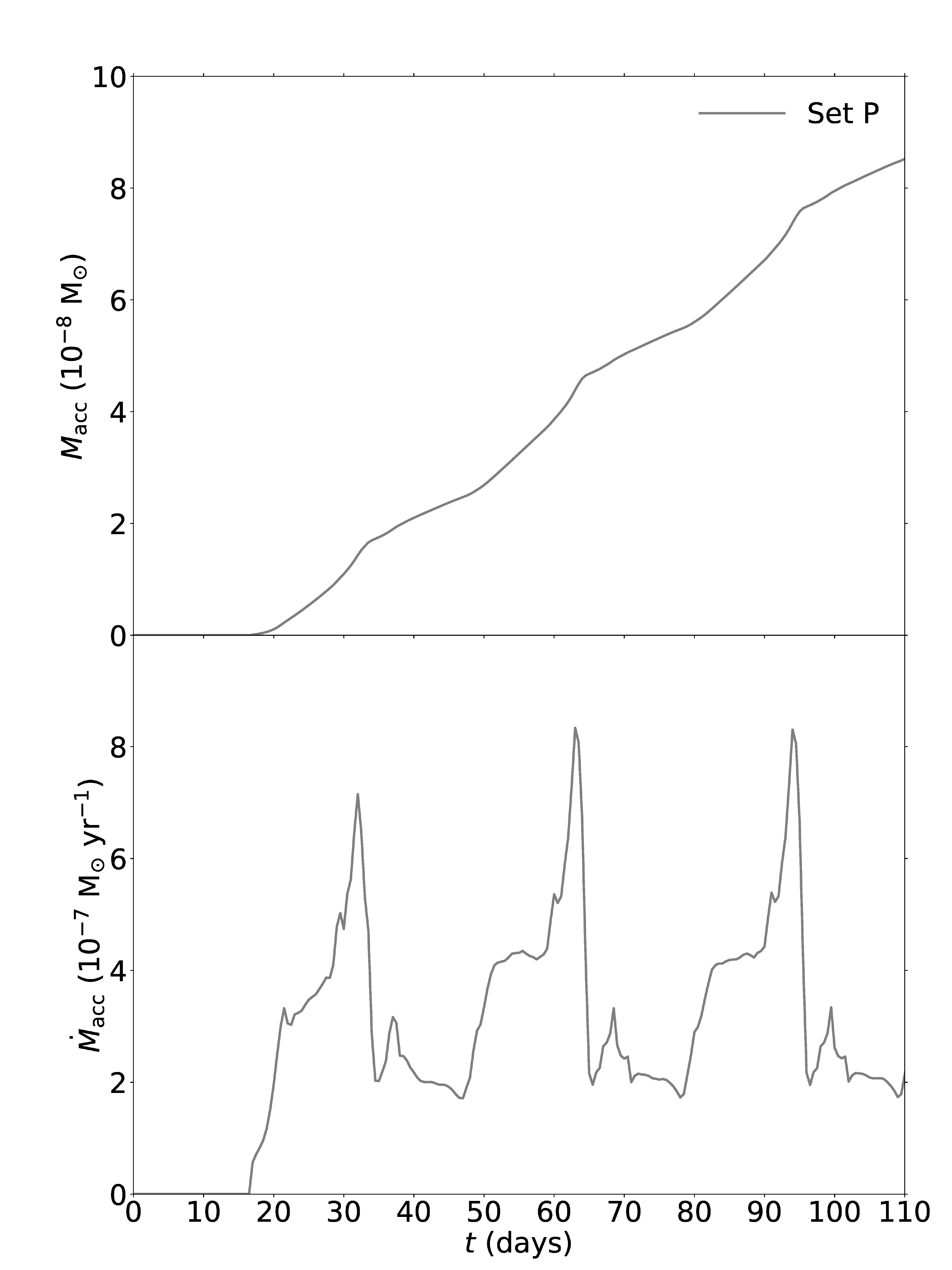} 
\caption{
The mass accretion rate obtained in our simulation or set P.
The first two orbits ($t<62 \days$) are the the initialization of the simulation, where the structure of the wind does not yet represent the system. We ran the simulation for seven orbitals period (the first three are shown in the figure) and from the third orbit we obtain repetitive behaviour of the accretion rate from orbit to orbit.
We can see the accretion rate is periodic with the orbit, with a peak just before periastron.
A sharp decline following the peak indicates the effect we term the Lizard Autotomy Effect, in which the primary cuts the accretion tail, and by that reduces the mass accretion rate.}
\label{fig:mdotP}
\end{figure}

Figure \ref{fig:mdotP} shows the accretion rate onto the secondary as a function of time.
The accretion we found is modulated by the orbital period. Accretion occurs throughout the binary period but much stronger close to periastron.
When the system is close to apastron there is still accretion taking place from residual material that the primary left around the secondary.
There is also a temporary formation of an accretion disc around the secondary that dissipates shortly after periastron passage. The residual accertion is from filaments of gas compressed by the orbital motion that remain gravitationaly bound to the secondary.

\begin{figure}
\centering
\includegraphics[trim= 1.2cm 0.0cm 0.0cm 0.0cm,clip=true,width=0.99\columnwidth]{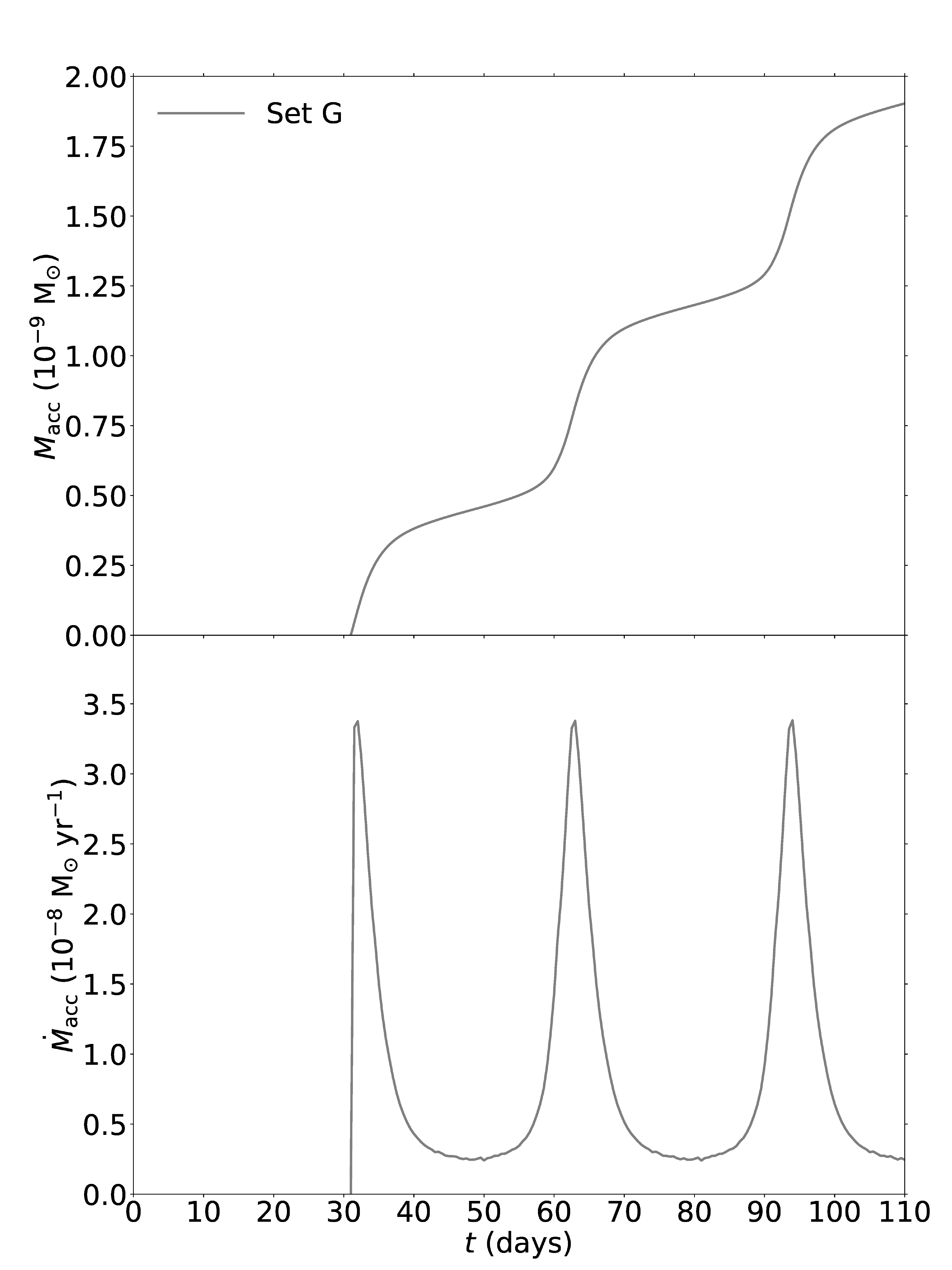} 
\caption{Similar to Figure \ref{fig:mdotP}, but for the G parameters set. The accretion rate onto the secondary is $\propto r^{-2}$. The Lizard Autotomy Effect is not evident. The mass accretion rate is much smaller that for set P.}
\label{fig:mdotG}
\end{figure}

\cite{Porteretal2021} suggested that the accretion of the primary wind by the secondary is the cause of the photometric variations at the orbital period in GG~Car.
The photometric variability according to their observations is $\Delta m_V \simeq 0.2 \rm{mag}$, which corresponds to a decrease of about 20\% in luminosity.
This conclusion was reached after excluding other options like extinction, scattering, and reflection by circumbinary material.
While accretion onto the companion can generally account for increase in brightening of a star or dimming of this light, we argue that in the case of GG~Car accretion by itself cannot cause the observed brightening.

Using the mass loss rate of the primary $\dot{M}_1=2.2\times10^{-6} \msyr$ from \cite{Porteretal2021}, the resulted highest mass accretion rate onto the secondary obtained by our simulations is $\dot{M}_{\rm 2,acc} \simeq 8\times 10^{-7} \msyr$.
The resulted accretion luminosity is
\begin{equation}
\begin{split}
L_{\rm 2,acc} &\simeq 25 \left(\frac{M_2}{7.2\rmModot}\right) \left(\frac{\dot{M}_{\rm 2,acc}}{8\times 10^{-7} \msyr}\right) \\
 & \times \left(\frac{R_2}{7.2\rmRodot}\right)^{-1} \rmLodot, 
\end{split}
\label{L2acc}
\end{equation}
which is a negligible fraction of the primary luminosity $ L_{\rm 2,acc}/L_1 \simeq 1.4\times10^{-4}$. Moreover, even if all the mass lost from the primary gets accreted onto the secondary it will still not be enough to explain an increase of $\simeq 20$\% in luminosity.
The other option, dimming of the secondary by the accreted material, is also not enough to account for the observed increase in luminosity, as the secondary luminosity is at most $4800\rmLodot$, only $<3\%$ of the primary luminosity, and even if obscured completely will not suffice.

We therefore suggest another explanation -- attenuation of the primary flux by the surrounding material.
We suggest that the luminosity does not periodically increase as a result of the accretion onto the companion, but rather periodically decreases as a result of absorption by the surrounding material, mainly in the tail.
The origin of the surrounding material is the wind of the primary itself. The gravity of the companion and the eccentric orbital motion creates an accretion structure with disc and tail, as shown in Figure \ref{fig:3d}.
We take the emission to originate at the surface of the primary star, and calculate the optical depth along the line of sight till reaching a boundary of the simulation box
\begin{equation}
\tau(t)= \int \kappa(t) \rho(t) \,dl.
\label{tau}
\end{equation}
We then calculate the attenuation of the observed magnitude by
\begin{equation}
M_V(t) = M_{V,0}+A_V(t)= M_{V,0}+1.086\tau(t),
\end{equation}
Where $M_{V,0}$ is the non-obscured magnitude of the star and $M_{V}$ is the observed magnitude through the absorbing material.

\begin{figure*}
\centering
\includegraphics[trim= 5.8cm 0.5cm 5.8cm 2.0cm,clip=true,width=0.89\textwidth]{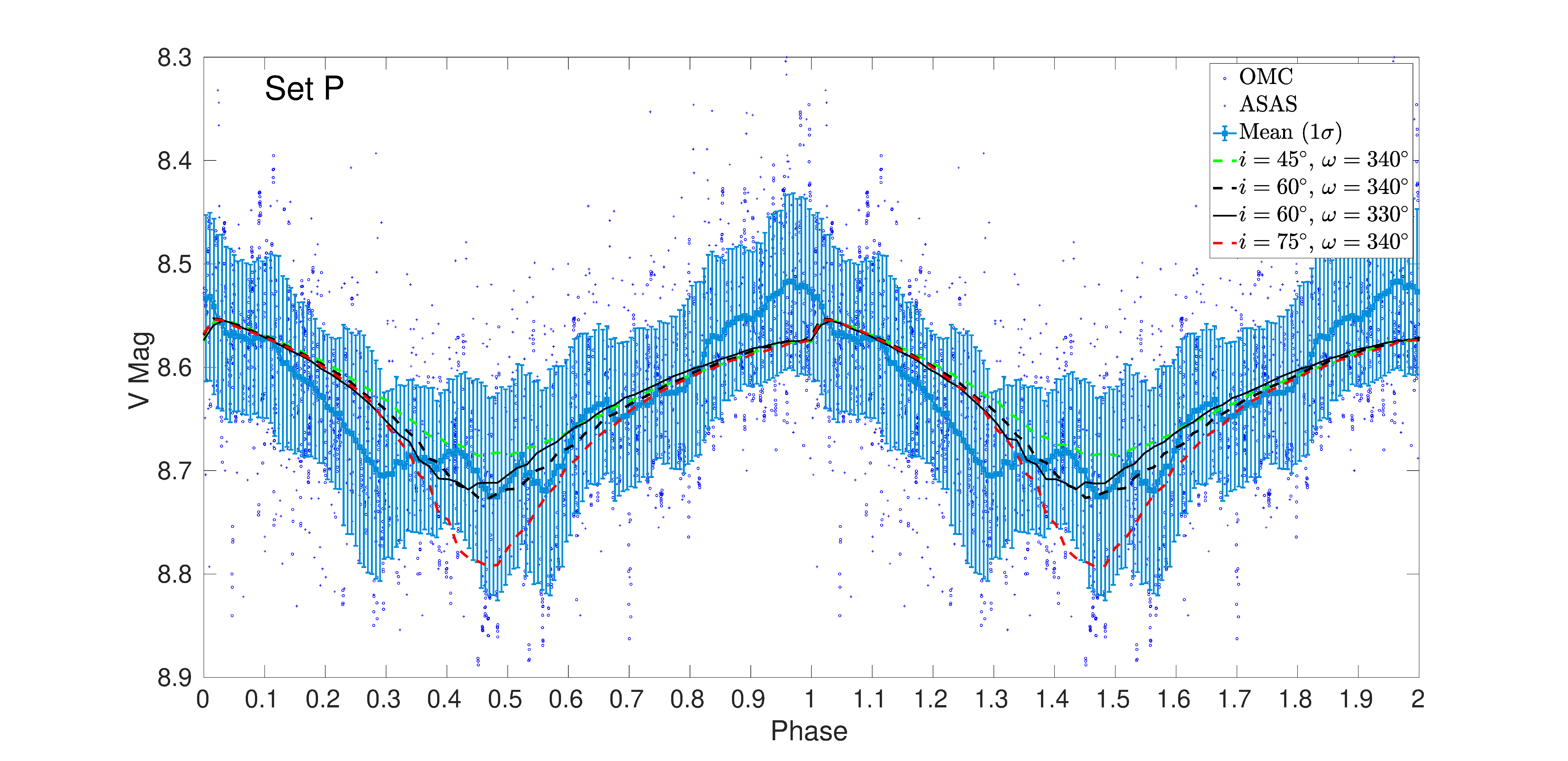} 
\caption{
The simulation results for the light curve of GG~Car for parameters set G.
The observed magnitude is taken from OMC (blue circles) and ASAS (blue crosses), and folded to the time of the orbit according to the ephemeris in Porter et al. (2021).
The observation results are binned into multiple bins, each includes 1/15 of the orbit, symmetric in both sides. For each bin the average and the standard deviation are calculated. The error bars indicate $1\sigma$.
The light curve follows the binary orbit. The tail and the spiral structure of the wind attenuate the light of the primary. The secondary is a negligible source.
The sharp rise close to periastron passage (phases 0, 1, 2) is the result of the primary cutting the tail and being exposed to the observer.
The resulted light curve from our simulation is given for different inclination angles and $\omega$. 
We find the orientation $i=60^{\circ}$ and $\omega=340^{\circ}$ gives the best results.
}
\label{fig:obscurationP}
\end{figure*}

The results of our simulation for set P on top of the observed magnitude is shown in Figure \ref{fig:obscurationP}.
The observed visible magnitude is taken from two sources and combined together, archival observations from the Optical Monitoring Camera (OMC) on-board the INTEGRAL satellite \citep{Domingoetal2010}, marked with blue circles and The All Sky Automated Survey (ASAS) catalog \citep{Pojmanskietal2003}, marked with blue crosses.
The observation results are binned into multiple bins, using a running window. Each bin includes 1/15 of the orbit, symmetric in both sides. For each bin the average and the standard deviation are calculated. The result is shown in blue squares with 1$\sigma$ error-bars.
The resulted light curve from our simulation is over plotted. We used four viewing angles with different inclination angles and $\omega$, within the range obtained by \cite{Porteretal2021} and \cite{GrantBlundell2022}.
We find the $i=60^{\circ}$ and $\omega=340^{\circ}$ gives the best results out of the four orientations we attempted, in agreement with \citep{Porteretal2021}.

The implication of our results is that the primary luminosity is in fact $\simeq 25\%$ larger than previously derived, namely $\log(L_1/\rmLodot) = 5.36 \pm 0.2$.
This also indicates that the mass of the primary of GG~Car is larger, and by extension also that of the secondary, as the derivation of the latter relies on the mass ratio between the two stars \cite{Porteretal2021}.

The tail and the spiral structure obtained in our simulations should also be evident in other observations. To show that, we process our simulation for set P to obtain the RV of the absorption component for the He I lines.
We assume that the lines originate on the surface of the primary.
For each length $dl$ along the line of sight to the primary we calculate $d\tau$ and the projected velocity of the gas along the path towards the observer $\mathbf{v_p}=(\mathbf{v} \cdot \mathbf{dl} ) d\hat{l}$, where $d\hat{l}$ points to the line of sight, namely $i=60^{\circ}$ and $\omega=340^{\circ}$.
We thus obtain $d\tau(v_p)$ along the line of sight, and get the radial velocity of the absorption. We repeat the process for every $0.5$ day along the orbital period.

\begin{figure}
\centering
\includegraphics[trim= 2.5cm 0.8cm 3.6cm 1.8cm,clip=true,width=1.0\columnwidth]{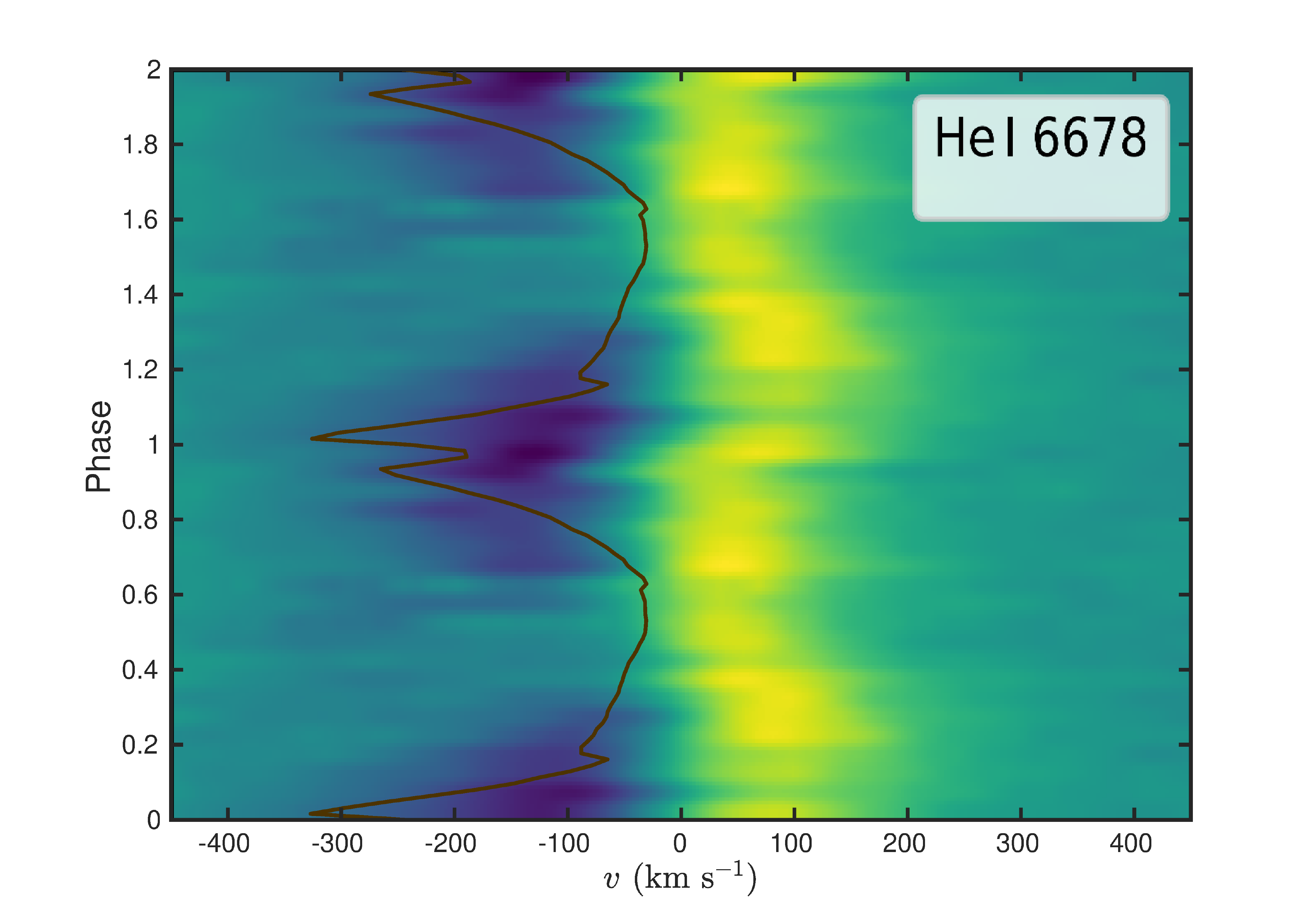}
\caption{
The result of our simulation for the RV of the absorped projected velocity, assuming line of sight in the direction $i=60^{\circ}$ and $\omega=340^{\circ}$ (brown line).
The results are plotted on top of the radial velocity profile of the He I 6678 line from Porter et al. (2021), taken by GJW (the color background with bluer colors represent absorption and yellower colors represent emission; precise colormaps were not provided).
The simulation results follow the observed absorption component, showing minimum values close to apastron (phase 0.5, 1.5) and maximal values close to periastron passage (phase 0, 1, 2).
}
\label{fig:RV_abs_HeI6678}
\end{figure}

\cite{Porteretal2021} showed three trailed spectra of He~I lines along the orbital period.
The absorption RV showed periodic variation, having larger values close to periastron.
In Figure \ref{fig:RV_abs_HeI6678} we plot our simulation result over figure 4 of \cite{Porteretal2021} that shows the trailed spectra of He~I 6678 taken by the Global Jet Watch (GJW) telescopes.
As we do not use advanced radiation transfer procedure a perfect match to a specific line is not expected.
However, our results show that the RV obtained from the simulation (brown line) generally follows the same behaviour, having minimal values close to apastron and maximal values close to periastron passage.

\subsection{Set G}
\label{sec:resultsG}

We repeat the analysis for parameters set G. Figure \ref{fig:dens_6panel_set_G} shows the density plots for set G.
For a large terminal velocity $v_{\rm 1,inf}=1400 \kms$ (set G) the orbital velocity at periastron is much smaller $v_{\rm orb,p}<<v_{\rm 1,inf}$. Therefore a curved tail is obtained but the primary does not cut as in set P and the Lizard Autotomy Effect is non existent. The result is a flow with a narrow curved tail behind the secondary, that changes relatively slowly along the orbit.
To have a similar Lizard Autotomy Effect as in set P, a much larger eccentricity is required.

Therefore for set G the mass accretion rate onto the secondary changes almost as $r^{-2}$ due to the orbital separation and no other effects are involved. The obtained accretion rate is shown in Figure \ref{fig:mdotG}. The peak accretion rate for set G is $\dot{M}_{\rm 2,acc} \simeq 3.4\times 10^{-8} \msyr$, much smaller than for set P.
The reason is the absence of the accretion tail in set G, that leads to no efficient focusing mechanism of material into the secondary.

\begin{figure*}
\centering
\includegraphics[trim= 5.8cm 0.5cm 5.8cm 2.0cm,clip=true,width=0.89\textwidth]{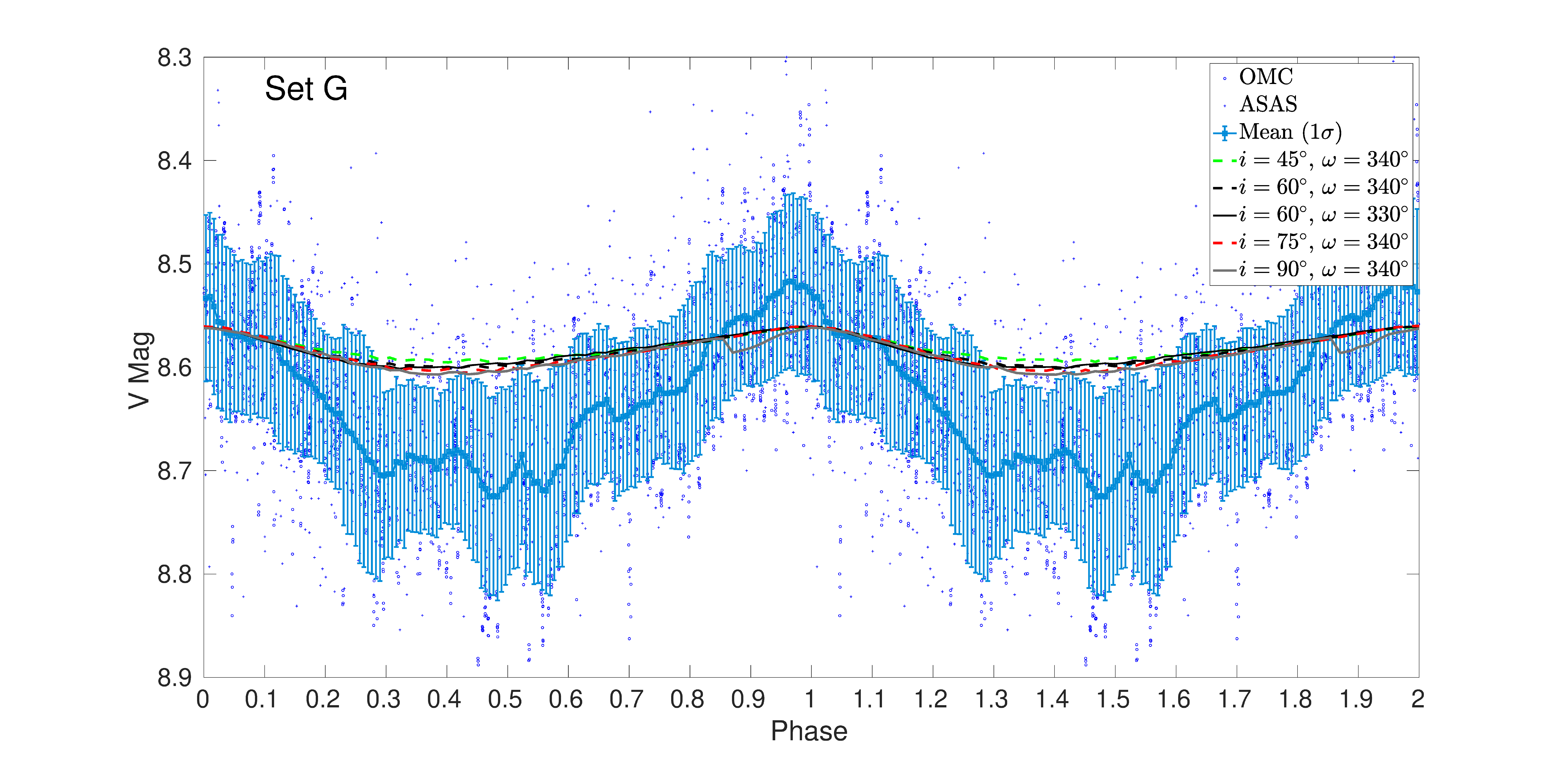} 
\caption{
The simulation results for the light curve of GG~Car based on parameters set G.
The observations are as in Figure \ref{fig:obscurationP}.
The resulted light curve from our simulation is given for different inclination angles and $\omega$.
The obscuration is periodic and follow the orbit.
For $i=90^\circ$ an eclipse is seen at phase $\simeq 0.87$.
We find that for none of the lines of sight the obscuration is sufficient to account for the decline in the light curve.
}
\label{fig:obscurationG}
\end{figure*}

In Figure \ref{fig:obscurationG} we calculate the obscuration from different lines of sight.
We obtain that the magnitude of variation is much smaller than the observed one, despite the mass loss rate of the primary being almost twice larger than in set P.
No result from the lines of sights that we checked (in any inclination) was found to reproduce the observed periodical light curve.
Even for $i=90^\circ$, where a maximal eclipse is obtained, the decline is not enough to account for the observed light curve. Moreover, even if we were to change the line of sight such that the eclipse is at phase 0.5 it would not be enough, as the observed decline is larger.

\section{DISCUSSION}
\label{sec:discussion}

Between the two sets of stellar and wind parameters that we tested, we found that the set of parameters we adopted from \cite{Porteretal2021} is a much better fit with the observations.
For this set we found that the primary cuts the accretion tail close to periastron passage -- a process we termed the Lizard Autotomy Effect.
A representation of this effect is shown in Figure \ref{fig:3d}.
The Lizard Autotomy Effect that we obtained in this paper causes the primary to self accrete mass from wind that it had blown earlier and formed the tail, and creates an expanding spiral shell with broken segments that expands outwards.

We suggest that such tails exists in other B[e]SGs systems and can be the source of the circumstellar material observed in such systems.
The accretion onto the secondary also forms a disc around the secondary (not a circumbinary disc) near periastron that later disperses. Some of the gas later remains, and form dense filament (as seen in Figure \ref{fig:dens_6panel}).  We suggest that the formation of such a disc can be accompanied by the launching of two opposite jets perpendicular to the plane of the disc. These two jets can push ambient gas and form a bipolar structure.

Such bipolar structures were observed around some B[e]SGs.
For example, \cite{Liimetsetal2022} showed an H$\alpha$ image of the nebula around MWC 314, a well known B[e]SG, that clearly shows its bipolar structure.
\cite{MarstonMcCollum2008} observed nebular features of the B[e] star MWC 819 seen in 2001, later to be confirmed as bipolar in a deeper image taken in 2020 by \cite{Liimetsetal2022}.
The B[e] star LHA~120-S~35 exhibits noticeable changes in its spectral patterns in both the optical and infrared regions and its P-Cygni line profiles of H~I, Fe~II, and O~I indicate the presence of a robust bipolar clumped wind \citep{Torresetal2018}.
The B[e]SG MWC 137 is surrounded by an optical nebula that was modeled as consisting of two double cones, though might be related to stellar pulsations \citep{Krausetal2017,Krausetal2021}.
\cite{SanchezContrerasetal2019} suggested an accretion disc around a compact object and jet-launching as a model for forming the red square nebula around the B[e] star MWC~922.

It is well established that jets can shape bipolar nebulae. The process applies to various astrophysical objects \citep{Frank1999} including planetary nebulae \citep[e.g.,][]{Soker1990,Soker1998,Soker2002,SokerLivio1994,Franketal1996,SahaiTrauger1998,AkashiSoker2008,SokerKashi2012,Balicketal2019},
post-AGB nebulae that involve a sudden anisotropic ejection of stellar gas by a very late AGB (or very early post-AGB) star during a common-envelope phase \citep[e.g.,][]{LeeSahai2009,Sahaietal2017,Soker2020, Zouetal2020, Bollenetal2022, Lopez-Camaraetal2022},
young stellar objects \citep[e.g.,][]{Shuetal1994, RayFerreira2021}
Luminous Blue Variables \citep[e.g.,][]{Notaetal1995,Soker2001,KashiSoker2010},
core collapse supernovae \citep[e.g.,][]{Gilkisetal2019,AkashiSoker2021, Soker2022,Akashietal2023},
and intermediate-luminosity optical transients \citep[e.g.,][]{SokerKashi2016}.
For planetary nebulae there are also indications that impulsive ejection of gas creates a bipolar structure \citep[e.g.,][]{AkashiSoker2008}, a similar idea to what we propose here for GG~Car and similar B[e]SGs.

The evidence for bipolar structure around B[e]SGs implies that very likely jets ejected from these stars were involved in their formation. This, in turn, implies there should exist an accretion disc to form these jet  that exists either continuously or intermittently.
\textit{Our simulation obtained such an accretion disc around the secondary} that exists for every orbit close to periastron passage, and therefore support the observed bipolar structure of nebulae around B[e]SGs.

As earlier noted, some B[e]SGs show large-scale nebulae and ejecta, suggesting previous mass-loss events \citep[e.g.,][]{Krausetal2021}.
These might be the result of the expanding spiral arms that dilute and merge with those of previous cycles.
The spectroscopic study of \cite{Porteretal2022} found that GG~Car has two circumbinary rings. These rings are located further out from the simulation grid. However, the outflow we find, in the shape of a spiral moving outward, could have formed such rings further out. This is likely if the spirals decelerate and accumulate in the radial direction, to form ring structures.

We suggest that the circumstellar disc around B[e] stars, or at least some B[e] stars, is in fact a cropped spiral created by the interaction of the stellar wind with the companion star.
The results here demonstrate that accretion is an important process in massive binaries, and should be considered in every close binary system that has uneven stellar wind momentum, and especially in ones with eccentric orbit.

\section{SUMMARY and CONCLUSIONS}
\label{sec:summary}

We used hydrodynamic simulations to show that the B[e]SG binary system GG~Car experiences accretion onto the secondary close to periastron passage.
By that we confirm the earlier suggestion for this process to occur in this system by \cite{Porteretal2021}.
We calculated the accretion rate onto the secondary star and found peaks in the accretion rate (Figure \ref{fig:mdotP}) as suggested by \cite{Porteretal2021}.
An accretion disc-like structure around the secondary is obtained shortly after periastron passage, and dissipates as the system approaches apastron (Figure \ref{fig:dens_6panel}). We also found that an asymmetric spiral structure is formed around the system.

We showed that the variation in the V light curve is not as a result of an increase to the luminosity from accretion, as the accretion power is not sufficient to account for it. Instead we propose that the primary is more luminous and it is dimmed periodically as the mass that it lost is compressed into the accretion tail and the spiral structure which adsorbs its light (Figure \ref{fig:obscurationP}). The implication is that the primary of GG~Car is more luminous by $\approx 25\%$ than previously derived, and consequently that the stellar masses are larger.

Accretion close to periastron passage is known to occur in other stars. One famous example is $\eta$~Car, an LBV star that shares similar properties with GG~Car. Close to periastron in the eccentric ($e=0.9$) and long ($P=5.54 \yrs$) orbit the star accretes mass from the equatorial region \citep{KashiSoker2008,Kashi2017}, and a disc is formed and stays for a few weeks, and accounts for multiple observations across the spectra \citep{KashiSoker2007a,KashiSoker2009b}, including the variation in He lines \citep{KashiSoker2007b}, as in GG~Car.

\cite{Porteretal2021} suggested that the fact that the blueshifted absorption component in GG~Car becomes deeper around periastron passage likely indicates that more mass is being lost from the primary’s wind at that time.
They concluded that the primary mass-loss rate and the rate of mass being captured by the secondary strongly depends on the phase of the binary due to the eccentric orbit, and therefore we can expect that there will be changes
in the brightness due to varying rates of mass transfer and accretion.
In our simulation we kept the mass loss rate of the primary fixed, and did not enhance it near periastron.
We agree with \cite{Porteretal2021} that the mass loss enhancement is reasonable.
However, the mass accretion rate is modulated according to the orbit regardless of the primary mass loss rate, and this is indeed obtained for the two sets of parameters we used. Enhancing the primary mass loss rate close to periaston can let alone enhance the accretion, but accretion exists even without the enhancement.

In the present simulation we did not modify the stellar atmosphere of either star as a result of accretion. Accretion at a very high rate can lead to a reduction in the effective temperature of the star and in turn weaken its wind, as suggested by \cite{KashiSoker2009b} for accretion close to the periastron passage of $\eta$~Carinae. This effect was later obtained in the simulations of \cite{Kashi2017}.  

In this paper we tried two representative sets of parameters. We did not make an attempt to run the simulation with tuning the parameters, such as stellar masses, radii, effective temperatures and eccentricity.
We showed that one of these parameter set (set P) reasonably reproduce photometric and spectroscopic observations.
The parameter set that reproduced the observations has a lower primary star wind velocity then expected from stellar evolution models, revealing an interesting problem that is beyond the scope of this work.

There are many spectroscopic observations for GG~Car, and our simulation and following analysis is limited and cannot reproduce all of them.
However, we showed that the RV absorption of He~I~6678 matches the results of our simulation (Figure \ref{fig:RV_abs_HeI6678}). This line is emitted from the primary or its wind, very close to the stellar surface \citep{Porteretal2021}, depends on the line of sight and less sensitive to other parameters.
Running more simulations on a set of parameters can very likely improve the constraints on the stellar and binary parameters.
Likewise, processing the results with radiation transfer code can give much better reproduction of spectral line observations than the relatively simple procedures we performed in this work.

Our simulations show that binary interaction can have a significant role in the formation of B[e]SGs and in shaping their circumstellar ans circumbinary environment. Hydrodynamic simulations are an important tool for analysing massive binary systems with complex wind structure, and in particular B[e]SGs, as done in this case.

\section*{Acknowledgements}
We thank an anonymous referee for helpful comments.
We acknowledge support from the R\&D Authority at Ariel University.
We acknowledge the Ariel HPC Center at Ariel University for providing computing resources that have contributed to the research results reported within this paper.

\section*{Data Availability}
OMC V-band photometry observations are available at \url{https://sdc.cab.inta-csic.es/omc/secure/form busqueda.jsp}.
\newline
ASAS V-band photometry observations are available at \url{http://www.astrouw.edu.pl/cgi-asas/asas cgi get data?105559-6023.5,asas3}.
\newline
This paper used Geneva code stellar evolution tracks from SYCLIST \url{https://www.unige.ch/sciences/astro/evolution/en/database/syclist/}.
\newline
The data underlying this article will be shared on a reasonable request from the corresponding author.

\label{lastpage}
\end{document}